# THz Emission from Spintronic Microstructure

ABBAS GHAFFARI,[1] MALEK ABDELSAMEI,[2] PUJA THAPA,[2] SEIJI MITA,[3] RAMÓN COLLAZO,[3] KENAN GUNDOGDU,[2] DALI SUN[2*] AND QING GU[1,2*]

[1]*Department of Electrical and Computer Engineering, North Carolina State University, Raleigh, NC 27695, USA*
[2]*Department of Physics, North Carolina State University, Raleigh, NC  27695, USA* [3]
[3]*Adroit Materials, Cary, NC 27518, USA*
[4] *Department of Materials Science and Engineering, North Carolina State University, Raleigh, NC 27695, USA*
*\* dsun4@ncsu.edu, qgu3@ncsu.edu*

**Recent advancements in spintronics have opened a new avenue in terahertz (THz) radiation sources that may outperform the traditional contact-based metallic counterparts. Inspired by the generation of broadband spintronic THz signals at the interface of a ferromagnet and ultrawide bandgap semiconductors, here we investigated the generation of THz radiation from micro-structured heterostructures of a metallic ferromagnet ($Ni_{80}Fe_{20}$) and an ultrawide bandgap semiconductor (AlGaN/GaN) that contains a layer of 2D electron gas. By precisely tailoring the dimension of the subwavelength pillars of a THz device, the micro-structured spintronic THz emitter can achieve up to more than three times higher emission intensity compared to that of the un-patterned counterpart. Our study advances the development of the next generation of spintronic THz sources that allow a tailored emission frequency and intensity control and, further, are compatible with existing integrated wide-bandgap semiconductor circuits.**

The THz frequency band of 0.1 – 20 THz possesses compelling optical properties due to its non-ionizing photon energy and unique spectral signatures for detecting large biomolecules, chemicals, and greenhouse gas emissions[1,2]. These properties enable a wide range of applications, such as non-invasive environmental monitoring and security examination, as well as high-speed wireless communication and computing beyond 5G technologies[3–5]. Additionally, the THz frequency range meets with the resonance frequencies of most low-energy quasi-particles, making THz spectroscopy a useful tool for investigating coherent light-matter interactions in solids[6].

One essential component of a THz system is a THz source. A common approach to generate THz radiation is to utilize the ultrafast charge of carriers through photoconductive antennas[7], nonlinear crystals[8], air plasma[9,10] and a new class of THz source – spintronic THz emitters[3–5,11–14]. Spintronic emitters can generate a gapless broadband THz spectrum, with a versatile choice of pump laser wavelengths, lightweight, minimal Joule heating, and low fabrication cost. Moreover, spintronic THz emitters have unique features that may outperform conventional THz emitters, such as active control of the phase of the THz pulse[14,15] and elliptical polarized emission by varying the magnitude/direction of the external magnetic field[16].

A spintronic THz emitter consists of a thin film of ferromagnetic material that is interfaced with a layer of heavy non-magnetic metals such as platinum (Pt)[3,4], hybrid organic-inorganic perovskites[5], 2D materials[17], and more recently, with wide bandgap semiconductors[14,15]. The development of spintronic THz emitters has gone through several key milestones[12], for example, by using non-magnetic metals with high electron mobility[12,18,19] and large spin-Hall conductivity[13]. Moreover, by tuning heterostructure thicknesses[13,20] and optimizing trilayer and multilayer structures[3,21], particularly those with opposing spin-Hall angles[4,13], the amplitude of the emitted THz radiation has been significantly boosted[4]. Also, Micro- and nano-patterning offers a powerful tool to engineering the bandwidth, peak frequency, direction, polarization state, and spatial mode of the THz emission[16,22–27]. For instance, ref [28] demonstrated that micro-patterning a Fe/Pt bilayer into stripe arrays allows the modulation of the emitted THz spectrum, including shifting the peak frequency and altering the bandwidth[28]. Ref [16] investigated spintronic-metasurface emitters with patterned Fe/Pt stacks, achieving simultaneous beam steering and polarization control of the THz signal by altering the magnitude and orientation of the external magnetic field. Ref [25] studied the coupling between a THz dipole in a spintronic emitter and an antenna and showed that a bow-tie antenna or split-ring resonator integrated with the spintronic emitter can enable spatial mapping and control of THz emission modes.

We recently demonstrated spintronic THz emission from ferromagnetic NiFe and n-doped GaN thin film heterostructures, here, we report the manipulation and enhancement of spintronic THz emission in a ferromagnet/UWBG semiconductor heterostructure, specifically, NiFe/AlGaN 2D electron gas/GaN, through micro-patterning the spintronic THz emitter. By directly patterning the heterostructure into 2D arrays of micropillars with judiciously chosen pillar side length, height, and periodicity, we can control the resonance frequency of the THz emitter through excitement and coupling of the cavity plasmonic mode with THz emission of the ferromagnet/UWBG heterostructure, thus allowing tailored emission enhancement at specific, desirable frequencies. Thanks to the bandgap tunability and integrated circuit (IC) compatibility of UWBG semiconductors, our micro-structured semiconductor-based spintronic THz emitter has a path toward integrating with existing electronic THz IC components.

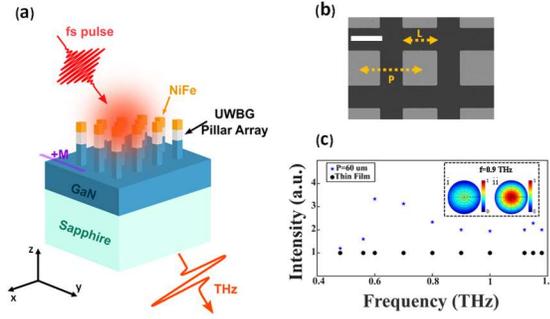

Fig. 2. Device structure and FDTD Simulation. (a) Schematic of THz generation from an array of micropillars spintronic THz emitter. The height and side length of the pillars is fixed at 500nm and 50 μm, respectively. (b) SEM image of a typical device. The scale bar is 50 μm. (c) Numerically simulated enhancement factor of the micro-patterned device. The y-axis represents the maximum intensity of THz emission, normalized to the maximum intensity of the unpatterned thin-film emitter. The inset displays the far-field radiation patterns at a frequency of 0.9 THz, where the left and right panels correspond to the unpatterned thin film and structured device, respectively.

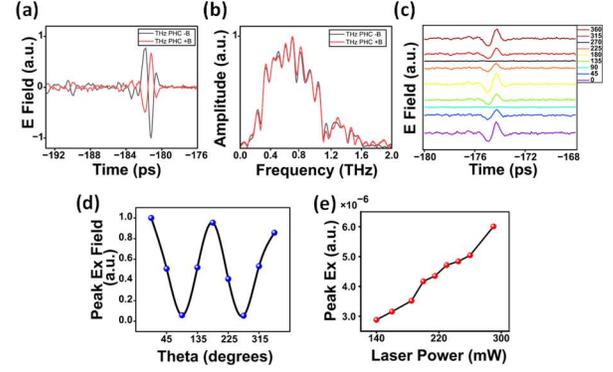

Fig. 1. Spintronic THz emission from THz micropatterned device. (a) The time trace of measured THz E-field from structured THz emitters under positive (red) and negative (black) external B-field. (b) The corresponding transmission spectrum of signals in A. (c) Time trace of THz pulses under different polarization angle of a wire-Grid polarizer. (d) The extracted cos(2θ) dependency. (e) Pump power variation shows a linearly polarized THz emission.

Our material platform is NiFe/AlGaN 2D electron gas/GaN on sapphire substrate. A 5 nm-thick film of NiFe as the ferromagnetic metallic layer is deposited on top of a 20 nm AlGaN on 1.2 μm-thick GaN on sapphire (see Supplementary 1 for more details). The vertical stack is patterned into a square lattice of micropillars using photolithography, and then etched into a array of THz micropillars with 20 periods, namely, 400 pillars arranged in a 20 × 20 matrix. Unless otherwise noted, the side length, L, of a single square pillar is 50 μm, and the height of the pillars, i.e., the total etched depth, is about 500 nm. The periodicity of the device, P, defined as the center-to-center separation of two pillars, varies from 60 μm to 130 μm. **Fig. 1(a)** schematically depicts the structure of our spintronic THz emitter, with a typical scanning electron microscope (SEM) image shown in **Fig. 1(b)**. Here, the generated THz E-field is a cross product of the spin diffusion direction ($J_S$) and spin polarization vector ($\sigma_S$) as described by **Equation1**:

$$E_{THz} = P_S \theta_{SH} (J_S \times \sigma_S) \quad \text{Equation 1}$$

where $P_S$ is the spin injection efficiency at the NiFe/AlGaN interface, which can be controlled by the bandgap alignment between the ferromagnet and the adjacent nonmagnetic material. The parameter $\theta_{SH}$ is the spin-to-charge conversion efficiency of the nonmagnetic layer, which is proportional to the strength of the spin-orbit coupling or Rashba parameter. $J_S$ and $\sigma_S$ are unit vectors of spin current density and spin polarization, respectively. A detailed investigation of the principle and mechanism of THz generation from ferromagnetic/semiconductor heterostructure was discussed in our previous work[15] with a brief summary provided in the Supplementary Information.

In order to investigate how micro-patterning influences the performance of THz emission, we designed and fabricated resonantly enhanced THz emitters. **Fig. 1(c)** illustrates FDTD simulation of THz emission, as a function of frequency for two configurations: one is a patterned structure with a periodicity P=60 μm (blue stars) and the other is an unpatterned thin film (black circles). The maximum intensity values are shown with respect to the unpatterned film that serves as a reference. These results highlight the collected signal enhancement in the patterned device. In these simulations, we position the THz dipoles at the interface between the ferromagnetic layer and the substrate. For the patterned structure, the dipoles are expected to be coupled to localized plasmon modes, in THz region, in the ferromagnetic layer on top of each pillar, resulting in resonant enhancement of the THz emission at specific frequencies. This resonant enhancement is more noticeable in the higher intensity peaks in the patterned structures, particularly in the 0.5 – 1.1 THz range. The inset depicts example far-field radiation patterns at 0.9 THz for both configurations. Radiation pattern (i), for the unpatterned thin film, shows a weaker radiation profile, while radiation pattern (ii), for the patterned structure, exhibits a more intense and focused radiation profile, supporting the hypothesis that plasmonic interactions in the ferromagnetic layer enhance the THz emission. This radiation pattern of the patterned structure demonstrates that coupling between the THz dipole sources and the plasmon modes of the heterostructure results in stronger, more spatially directed THz emission (**Fig. S1**). A second example at 1.2 THz (**Fig. S2**) further supports this conclusion: the patterned structure displays a highly asymmetric directional radiation profile, indicative of directional emission influenced by plasmonic mode coupling, whereas the reference device exhibits a nearly symmetric radiation pattern, consistent with isotropic dipole emission and the absence of plasmonic enhancement. Note that due to the lack of optical properties of NiFe in the literature, we used those of Ni in the simulations (see Supplementary 1 for more details).

Our THz emitters are studied using a homemade time-resolved THz spectroscopy setup, as described in Supplementary 1, where an 800 nm femtosecond laser pulse excites NiFe and creates hot electrons. The unpatterned thin-film version of the same material stack is used as a control sample. To confirm the spintronic nature of the THz emission, we measure the emitted signal under different configurations of the external B-field. **Fig. 2(a)** depict the time domain THz E-field signals of the micropatterned samples (bottom curves) for both positive and negative external B-fields. The phase of the THz pulse reverses upon switching the direction of the applied B-field. This observation strongly supports the spintronic origin of the THz emission because rotating the magnetization of the ferromagnetic layer by 180 degrees will flip the spin orientation of the hot electrons, thereby altering the direction of the transient charge current and the resulting E-field[15]. **Fig. 2(b)** shows the corresponding frequency domain signal of the oscillating transient charge current of **Fig. 2(a)**, by taking the Fourier transform of time domain signal. The bandwidth of either spectrum extends up to more than 1.5 THz – the frequency limit of our ZnTe detector. It is important to note that the oscillatory feature in the spectrum arises from water absorption. To measure the polarization state of the emitted THz radiation, a wire-grid polarizer is placed in the beam path between the THz emitter and the detector, with its polarization angle, θ, varied from 0° to 360°. The typical THz signal at different polarizer angles is shown in **Fig. 2(c)** We observed that the THz pulse is strongest at θ =0° and 180°, while the weakest at θ = 90° and 270°. The cos(2θ) dependency shown in **Fig. 2(d)** indicates the linear polarization of the THz radiation, which is perpendicular to the spin current and the applied B-field, Thus, our micro-patterning of the heterostructure does not alter the polarization state of the spintronic THz emission. To investigate how the excitation power influences emission behavior, we illuminate the device with a range of pump powers and monitor the output E-field pulse in the time domain. **Fig. 2(e)** illustrates the dependency of the THz E-field peak amplitude on the incident femtosecond laser power, which increases linearly with rising excitation power.

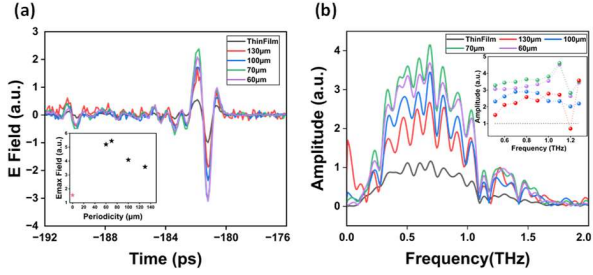

Fig. 3. Impact of micro-patterning on the emitter performance. (a) THz waveforms of micropatterned spintronic emitters. Different colors represent different periodicities with a fixed pillar size of 50um. The black curve indicates thin film emitter as the reference. The inset displays the peak-to-peak amplitudes, and the red star presents the thin film emitter. The data normalized so that the effective region deposited with NiFe is considered. (b) Corresponding amplitude spectrum, obtained by Fourier Transform of signals shown in (A). The inset demonstrates the amplitude ratio of patterned devices to the reference at selected frequencies.

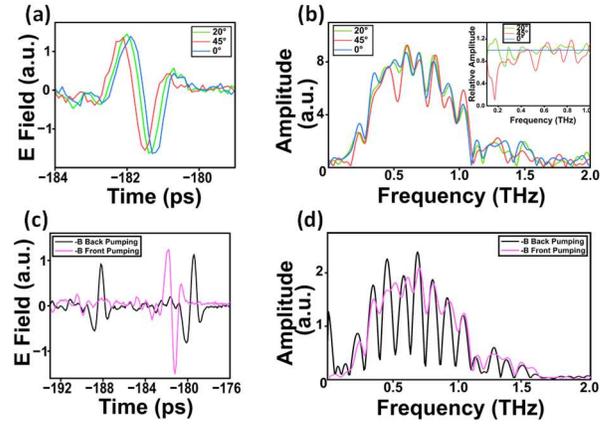

Fig. 4. (a) THz waveforms of micropatterned spintronic emitters under different excitation incident angles. (b) Corresponding amplitude spectrum, which is the Fourier Transform of signals shown in A. The inset demonstrates the amplitude ratio of inclined pumping to the normal incidence pumping (blue horizontal line). (c) THz emission under backside pumping from the sapphire substrate side. (d) Corresponding frequency-domain spectrum from signals in (c).

We demonstrated that our proposed micro-structured device generates spin-based THz emission. Now, we study the lattice period influence on the THz emitter performance. **Fig. 3(a)** presents the time-domain THz signal from devices with various periodicities, as well as from a thin film control sample. To eliminate the influence of the partial NiFe removal on the micropatterned samples (Due to etching), all signals are normalized by the laser spot area to account for the effective area covered by NiFe and excited by the femtosecond laser. A time-domain THz signal for the device of 60 μm and 70 μm periodicities shows a peak-to-dip amplitude that is more than 3 times larger than the thin film counterpart, highlighting the significant emission enhancement due to micropatterning with subwavelength pillars. The inset of **Fig. 3(a)** shows the maximum (peak-to-dip) E-field evolution with periodicity, further emphasizing that the 60 μm and 70 μm periodicity devices outperform the others. The red star in the inset presents the thin film emitter performance. **Fig. 3(b)** presents the Fourier spectra of the time-domain signal of **Fig. 3(a)**. While devices with 60 μm and 70 μm periodicities show the highest emission amplitude, as the periodicity increases above 100 μm, the emission amplitude significantly decreases, despite still being higher than the thin film control sample.

The inset of **Fig. 3(b)** shows the ratio of the THz signal from patterned devices of various periodicities to that of the thin film, plotted as a function of frequency, thus allowing us to quantitatively assess the emission enhancement from the microstructure devices. For example, for the 60 μm and 70 μm periodicities, the enhancement factors at 0.5 THz reach approximately 3 and 3.1, respectively, indicating a significant increase in THz emission efficiency. As the periodicity increases, the enhancement factor declines, with the 130 μm periodicity device showing an enhancement factor of only 1.6. Notably, enhancement is observed in the raw data (**Fig. S5**), even without normalization, for devices with periodicities of 60 μm and 70 μm. Please note that the oscillatory behavior from water absorption complicates the identification of spectral changes or shifts due to patterning. While a comprehensive investigation of the enhancement mechanism is beyond the scope of this work, the normalized maximum intensity (enhancement factor) at similar frequencies for the patterned device with periodicity of 60 μm observed in (**Fig. 1C**) suggests a significant role of the plasmonic coupling in the enhancement. These results demonstrate the superior performance of the micropatterned devices, clearly indicating the effectiveness of introducing periodic structures in enhancing THz emission.

To study the response of the emitter to different excitation angles, we vary the incident angle of the femtosecond laser pump by rotating the device around y-axis. **Fig. 4(a)** shows that the time-domain signal remains consistent across different incident angles of

0°, 20°, and 45°, but there are slight variations in amplitude and a shift in the arrival time. At normal incidence, the peak of the THz pulse arrives slightly earlier compared to the pulses at 20° and 45°. This temporal shift suggests that the angle of incidence affects the path length and, consequently, the timing of the emitted THz pulses. **Fig. 4(b)** shows the corresponding frequency-domain signals, with the inset highlighting the amplitude ratios by comparing the amplitude at 20° and 45° to that at normal incidence (0°). Lastly, **Fig. 4(c)** shows the influence of pumping direction on the micropatterned emitter with the device of 70 μm period. When pumped from the substrate side, two distinct peaks are present. The first peak corresponds to the initial spin-to-charge conversion, which occurs immediately after the device is excited with a femtosecond laser pulse. The second peak, which is delayed by 8.8 picosecond relative to the first peak, arises due to the reflection of the spin current at the air/NiFe interface. Compared to front pumping, the spin current is inverted and subsequently, the phase of the THz waveform is flipped when excited from the backside, i.e., substrate. **Fig. 4(d)** depicts the corresponding frequency domain signal, showing variations in the amplitude of the THz spectrum. These results highlight the impact of pumping direction on the THz emission, emphasizing the role of spin current reflection in the observed signal.

In this work, we have demonstrated micropatterned spintronic THz emitters on a semiconductor platform. We show that significant enhancement of THz emission from their thin film counterparts can be achieved through patterning ferromagnetic/UWBG semiconductor heterostructures. We reveal that these periodic THz emitters, particularly those with periodicities of 60-70 μm, exhibit more than a three-fold increase in THz emission amplitude compared to unpatterned thin films. This enhancement correlates with the excitation of plasmons. Furthermore, our investigations into the effects of the excitation angle and pumping direction reveal additional ways to modulate and optimize the THz emission. Lastly, although we use the sapphire substrate in this work, the substrate can be readily adapted to GaN for better IC compatibility. These results demonstrate that these spintronic THz emitters can emerge as the next-generation broadband THz source with high efficiency and tailored emission characteristics.

**Funding.** Q. Gu acknowledges support from the National Science Foundation (ECCS-2240448 and ExpandQISE-2329027); D.S. acknowledges financial support from the Department of Energy under award number DE-SC0020992; K. Gundogdu acknowledges support from the Department of Energy under award number DE-SC0024396; R. Collazo acknowledges support from the National Science Foundation (ECCS-1916800) and the Army Research Office (W911NF-22-2-0171).

**Disclosures**. The authors declare no conflicts of interest.

**Data Availability Statement (DAS).** Data underlying the results presented in this paper are not publicly available at this time but may be obtained from the authors upon reasonable request.

**Supplemental Document.** See Supplement 1 for supporting content.